\documentclass[12pt]{article}
\usepackage{natbib} 
\usepackage{url} 
\usepackage{amsmath, amssymb, amsthm}
\usepackage{geometry}
\usepackage{booktabs}
\usepackage{array}
\usepackage{graphicx}
\usepackage{algorithm}
\usepackage{algpseudocode}

\newcommand{\blind}{0}

\begin{document}

\bibliographystyle{abbrvnat} 

\def\spacingset#1{\renewcommand{\baselinestretch}%
{#1}\small\normalsize} \spacingset{1}


\if0\blind
{
  \title{\bf Physics-Informed Basis Functions for Nonlinear Response
Curve Decomposition: A Parsimonious Alternative to Splines}
  \author{\hspace{.2cm}M. Ross Kunz\thanks{
    Corresponding author:  matthew.kunz@inl.gov}, 
     Jieun Lee, Jaden Palmer
    \\
    Idaho National Laboratory}
  \maketitle
} \fi

\if1\blind
{
  \bigskip
  \bigskip
  \bigskip
  \begin{center}
    {\LARGE\bf Title}
\end{center}
  \medskip
} \fi

\bigskip
\begin{abstract}
Curve fitting for physical, biological, and engineering data typically forces a choice between interpretable but rigid parametric forms and flexible but physically opaque smoothers.
This paper introduces the Growth-Decay Curve (GDC), a physics-informed basis derived as the product of a lognormal growth cumulative distribution function and an exponential decay term, each traceable to a governing differential equation.
GDC's parameters correspond directly to growth and decay timescales, yielding dimensionless ratios and shape descriptors that connect the fit back to the underlying dynamics.
Across simulated differential-equation solutions and applications spanning physical, biological, and economic data, GDC matches the fit quality of splines, generalized additive models, radial basis function networks, and Fourier regression, while remaining substantially more parsimonious and physically interpretable.
\end{abstract}

\noindent%
\vfill

\newpage
\section{Introduction}
\label{sec:intro}

A central challenge in the analysis of experimental data is determining the governing dynamics that produce an observed curve.
Physical processes are often first characterized through the differential equations governing their generative mechanism, with the resulting distributional form only later adopted as a general-purpose empirical tool detached from that original derivation.
Physics-based modeling (deductive reasoning via differential equations) and statistical modeling (inductive reasoning via machine learning) are closely related in this sense, yet are typically pursued separately in practice. 
To merge these two approaches, the objective is to propose a small, specific set of physically motivated basis functions, derived from a small number of canonical governing equations, to connect the physics to data through standard statistical estimation.

A range of existing approaches sits between the fully physics-based and fully data-driven extremes described above, and each informs some part of the basis proposed here.
The Fourier transform illustrates this range directly. Regardless of its origin in the heat equation, it is used today as a general-purpose decomposition of a signal into its dominant frequencies via the Fast Fourier Transform (FFT) algorithm, without reference to any specific governing equation that produced the signal \citep{baron1822theorie, robinson1982historical}.
Gaussian process regression (GPR) has been used extensively as a surrogate model for physics-based simulations, replacing an expensive forward solve with a flexible, uncertainty-quantified statistical emulator \citep{williams1995gaussian, gramacy2020surrogates, williams2006gaussian}.
Closely related, radial basis function (RBF) methods can be viewed as representing the posterior mean of a corresponding GPR under a matched kernel \citep{williams2006gaussian}. 
Separately, RBF methods have been shown to provide direct insight into physics-based processes through the Kansa method, which solves partial differential equations via RBF collocation \citep{broomhead1988multivariable, gorbachenko2023physics, stenkin2024mathematical, kansa1990multiquadrics, fasshauer1996solving, hon2001unsymmetric}.
Nonparametric regression, via splines and Generalized Additive Models (GAMs), is the standard approach for fitting a smooth response curve when its functional form is unknown \citep{hastie1986generalized,wood2025generalized}.
GAMs enforce structure through the choice of basis and link function, but that functional space does not directly correspond to any particular governing equation; they yield smooth, flexible estimates at the cost of physically interpretable parameters.
Standard spline fitting similarly does not draw on this kind of prior knowledge. The basis functions, such as polynomial pieces, B-splines, or truncated power bases, are chosen for mathematical convenience, continuity, compact support, or numerical stability, rather than physical relevance. The resulting knot locations and spline coefficients have no direct connection to diffusion coefficients, activation energies, or rate constants, and the smoothness penalty enforces regularity without encoding any expectation about the specific functional form implied by a governing equation \citep{wahba1990spline}.
This is not a limitation unique to splines, and related efforts have used spline or basis-expansion methods as a route to recovering governing equations directly from data \citep{sun2022bayesian,golder2025discovering, brunton2016discovering}, suggesting that the gap between physically motivated and mathematically convenient bases is narrower in practice.

Motivated by this gap, this paper proposes replacing a purely mathematically convenient basis with a physics-informed basis: a small set of functional forms derived as exact or asymptotic solutions to a limited set of canonical governing equations.
The proposed basis combines two distinct physical phenomena common to a wide range of rise-and-decay processes, a cumulative growth phase and an exponential decay phase, into a single functional form.
This construction connects data-driven curve fitting and physics-based differential-equation modeling within a single estimation procedure, preserving the physical interpretability of the fitted parameters while retaining much of the flexibility associated with nonparametric smoothing, and without requiring the analyst to specify or solve the governing equation in advance.

The remainder of the paper is outlined as follows. 
Section 2 describes how to construct a basis set from proxies of differential equations. 
Section 3 goes through the computational aspects of curve fitting with the new basis and variable projection.
Section 4 uses differential equations based simulations to estimate known key parameters within reaction kinetics and epidemic incidence.
Section 5 applies the methodology to measured data that has a common application in differential equations: economic response data, solar activity, and other rise-and-decay processes.
Finally, Section 6 provides a brief conclusion and discussion on the use of the developed methodology and future work that is not contained in this manuscript.

\section{Methodology}
\label{sec:basis}

This section builds the connection and intuition from canonical differential equations, e.g., exponential, diffusion, and drift-diffusion, to a basis set that provides a curve-fitting approach to experimental data. The objective is to provide a bridge from differential equations to data-driven techniques while maintaining interpretability.

Consider the problem of curve fitting, approximating a smooth function to represent a measured response $y$ given a single covariate $x$, i.e.,
\begin{equation*}
    y = f(x) + \epsilon,
\end{equation*}
where $\epsilon$ is assumed to be Gaussian noise with mean zero, independent of $x$.
Without loss of generality, assume $x \in \mathbb{R}^n$ is strictly increasing, i.e., $x_i < x_{i+1}$ for $i = 1, \ldots, n-1$, where $n$ is the number of observations.
Estimating $f(\cdot)$ is a classical problem, and a range of techniques, whether physics-based or statistical, can be used to obtain the best empirical fit.
These methods are distinguished by the complexity of the resulting model, e.g., its effective degrees of freedom, and by their interpretability, e.g., how transparently the fitted values can be traced back to the data.

A basis set built directly from physical governing equations offers a route to this balance, but at a cost.
Restricting each basis element to a function with a known physical origin necessarily narrows the space of curves the basis can represent.
The remainder of this section develops a basis set that accepts this trade-off deliberately.

\subsection{Three Canonical Differential Equations and Their Distributional Solutions}
\label{sec:basis_simple}

A wide range of physical processes can be described by three functional forms, each arising as the solution to a canonical differential equation and each encoding a distinct physical mechanism.
The simplest is exponential decay or growth, which emerges as the solution to
\begin{equation}
    \frac{dx}{dt} = kx, \qquad x(t) = x_0 e^{kt},
    \label{eq:exp-decay}
\end{equation}
where $x(t)$ is the state variable, or quantity of interest, at time $t$, $x_0$ is its initial value at $t=0$, and $k$ is the rate constant governing exponential growth ($k>0$) or decay ($k<0$).
When the quantity of interest spreads through space or state rather than simply growing or decaying in time, the governing equation becomes the diffusion (or heat) equation,
\begin{equation*}
    \frac{\partial y}{\partial t} = D \frac{\partial^2 y}{\partial x^2},
\end{equation*}
where $y(x,t)$ is the concentration (or density) of the diffusing quantity at spatial position $x$ and time $t$, and $D$ is the diffusivity governing how quickly the profile spreads.
The solution to this equation is the Gaussian, i.e., Normal distribution \citep{crank1979mathematics}.
The Gaussian, however, carries an implicit assumption: the domain is unbounded, the underlying process is additive, and the profile is symmetric about its mean.
Allowing the rate constant in \eqref{eq:exp-decay} to fluctuate stochastically, $k \to k + \sigma\,\xi(t)$, gives geometric Brownian motion,
\begin{equation}
    dx = kx\,dt + \sigma x\,dW_t,
    \label{eq:gbm}
\end{equation}
where $x$ is again the state variable, $k$ is the drift (or mean growth rate), $\sigma$ is the volatility, governing the size of random fluctuations in the rate, and $W_t$ is a standard Wiener process, representing the stochastic driving noise.
This stochastic differential equation (SDE) has a corresponding Fokker--Planck equation:
\begin{equation}
    \frac{\partial p}{\partial t} = -\frac{\partial}{\partial x}\big[kx\,p\big] + \frac{1}{2}\frac{\partial^2}{\partial x^2}\big[\sigma^2 x^2 p\big],
    \label{eq:fpe-gbm}
\end{equation}
where $p(x,t)$ is the probability density of the state variable $x$ at time $t$, and $k$ and $\sigma$ retain their meaning from \eqref{eq:gbm} as the drift and volatility of the underlying process.
Because \eqref{eq:gbm} is multiplicative through $x$ rather than additive, the process naturally enforces nonnegativity of the state variable, a constraint the Gaussian cannot impose.
The standard log-substitution $y = \ln x$, together with a shift to the drift-comoving frame $z = y - \mu t$ where $\mu \equiv k - \tfrac{1}{2}\sigma^2$, reduces \eqref{eq:fpe-gbm} to the heat equation,
\begin{equation}
    \frac{\partial p}{\partial t} = \frac{\sigma^2}{2}\frac{\partial^2 p}{\partial z^2},
    \label{eq:heat-eq}
\end{equation}
where $p$ is now the probability density expressed in the transformed, drift-comoving coordinate $z$, and $\sigma^2/2$ plays the role of the diffusivity $D$ from the heat equation above.
Solving \eqref{eq:heat-eq} and transforming back through $z \to y \to x$ recovers the lognormal transition density of \eqref{eq:gbm}, the distribution of the state variable $x$ at a fixed observation time $t$.
This log-substitution and change-of-frame reduction of the geometric Brownian motion Fokker--Planck equation to the heat equation is a standard result in stochastic calculus \citep{oksendal2003stochastic, risken1989fokker} and constrained-transport settings \citep{andersson2021mechanisms}.
Together, these three functionals (exponential, Gaussian, and lognormal) form a compact, physically motivated basis capable of representing a wide class of growth, transport, and decay profiles encountered in practice.

Physical systems often exhibit a combined growth-decay response arising from two sequential processes: a diffusion-limited arrival phase followed by an exponential departure phase.
On an unbounded, symmetric domain, the convolution of an exponential and a Gaussian distribution may describe this process. This convolution is commonly referred to as the Exponentially Modified Gaussian (EMG) distribution \citep{grushka1972characterization},
\begin{equation}
    \mathrm{EMG}(x;\, \mu, \sigma, \lambda) =
    \frac{\lambda}{2}
    \exp\!\left(\frac{\lambda}{2}(2\mu + \lambda\sigma^2 - 2x)
    \right)
    \operatorname{erfc}\!\left(
    \frac{\mu + \lambda\sigma^2 - x}{\sqrt{2}\,\sigma}\right),
    \label{eq:emg}
\end{equation}
where $x$ is the state variable, $\mu$ and $\sigma$ are the mean and standard deviation of the underlying Gaussian component, and $\lambda>0$ is the rate of the underlying exponential component.
The EMG is well-suited to symmetric, unbounded domains; however, as discussed above, it cannot enforce the nonnegativity required in many physical systems.

\subsection{The Growth-Decay Curve (GDC)}
\label{sec:gdc}

When the state space is strictly positive, underlying processes are often multiplicative rather than additive, and the convolution of an exponential with a lognormal distribution has no closed form.
Rather than approximate this convolution numerically, we define the GDC directly as the product of the lognormal CDF growth term derived above and an exponential decay term,
\begin{equation}
    \mathrm{GDC}(t;\, \mu, \sigma, \lambda) =
    \Phi\!\left(\frac{\log t - \mu}{\sigma}\right) e^{-t / \lambda},
    \label{eq:gdc}
\end{equation}
where $\Phi(\cdot)$ is the standard normal CDF, $\mu$ and $\sigma$ are the log-scale location and scale parameters of the growth term, and $\lambda > 0$ is the exponential decay timescale. 

Equation~\ref{eq:gdc} captures two physically distinct phases: the growth phase (diffusion-limited arrival) and decay phase (reaction or departure) in a single three-parameter functional form.
When a signal consists of multiple processes, a GDC mixture is the nonnegative combination of $m$ such components,
\begin{equation}
    f(t) = \sum_{i=1}^{m} \beta_i\, \mathrm{GDC}(t;\, \mu_i, \sigma_i, \lambda_i),
    \label{eq:gdc-mixture}
\end{equation}
where $\beta_i$ is the amplitude, or mixing weight, of the $i$-th component, each contributing its own growth phase and decay phase, with the amplitudes $\beta_i$ estimated jointly with the nonlinear shape parameters as described in Section~\ref{sec:computation}. Since each GDC component's parameters are related to variables of physical process via differential equations, fitting a GDC mixture to observed data is simultaneously a curve-fitting exercise and an estimation problem for those physical quantities.
This is what allows the basis set constructed above to be related back to the original differential equations using purely data-driven, statistical techniques, without requiring the governing equation to be solved or even fully specified in advance. 

\subsubsection{Parameter Interpretation}
\label{sec:gdcParameters}

The decay term $\lambda$ offers the most direct case.
In \eqref{eq:exp-decay}, $k$ is a rate with units of inverse time, and the corresponding characteristic timescale is $1/k$.
The decay term plays the same physical role, but is parameterized directly by the timescale itself rather than by a rate, i.e., the factor $e^{-t/\lambda}$ in \eqref{eq:gdc} uses $\lambda$ as a characteristic decay timescale.
The location parameter $\mu$ of the lognormal CDF factor plays the analogous role for the growth phase.
By construction, $\mu$ is the log-median of the underlying lognormal distribution, so that $\exp(\mu)$ is the median arrival time or median of a given autocorrelated process, e.g., spectral wavelength.
As such, the growth and decay can be compared as a single dimensionless ratio, $\Pi_\tau = e^{\mu} / \lambda$. 
In words, the growth timescale relative to the decay timescale, is a generic ratio of rates, structurally analogous to a ratio of characteristic timescales, e.g., the Damk\"{o}hler number in reaction-transport systems and the reproduction number $R_0$ in epidemiology \citep{damkohler1936einflusse, kermack1927contribution, diekmann1990definition}.
Because this timescale-ratio interpretation depends only on $\mu$ and $\lambda$ having units of time, and not on any assumption specific to a governing process, the same construction transfers directly to other domains in which a process can be described as a competition between a characteristic growth timescale and a characteristic decay timescale, expressed as a dimensionless number.

The remaining nonlinear parameter, $\sigma$, describes the shape or heterogeneity of the growth process itself.
As $\sigma \to 0$, the lognormal CDF factor $\Phi((\log t - \mu)/\sigma)$ collapses to a Heaviside step function centered at $t = \exp(\mu)$, so that the GDC reduces to a pure exponential decay switched on instantaneously at the growth timescale. 
As $\sigma$ increases in value, the onset of decay is smeared out over a longer, increasingly asymmetric window.
A complementary interpretation of $\sigma$ comes from comparing the coefficient of variation (CV) of the growth and decay components directly.
The exponential distribution has a CV that is equal to one, regardless of the value of its timescale $\lambda$, while the lognormal distribution's CV depends only on its shape parameter,
\begin{equation}
    \mathrm{CV}_{\ln} = \sqrt{\exp(\sigma^2) - 1},
    \label{eq:lognormal-cv}
\end{equation}
and is independent of $\mu$.
Since the exponential factor's CV is a fixed constant, comparing $\mathrm{CV}_{\ln}$ against one gives a second, shape-based comparison between the growth and decay phases of the GDC, complementary to the timescale ratio $\Pi_\tau$ discussed above. When $\mathrm{CV}_{\ln} > 1$, the growth phase is more dispersed than the exponential decay phase. When $\mathrm{CV}_{\ln}$  is less than one, the growth phase is comparatively more concentrated than the decay phase.

\section{Computation}
\label{sec:computation}

Fitting a mixture of $J$ GDC components to an observed response $y(t)$ can be viewed as fitting a mixture-of-densities model, but to the empirical curve itself rather than to a sample of realizations drawn from it.
In a standard mixture model, component parameters are estimated from individual observations, typically via the expectation-maximization algorithm.
Here, the ``observations'' are instead the values of the normalized response curve $\tilde{y}(t)$ evaluated over its domain, so the fitting procedure estimates each component's nonlinear shape parameters and linear mixing weight directly from that curve rather than from resampled draws.

The response is first normalized to a nonnegative curve with unit area over a unit-interval domain:
\begin{equation}
    \tilde{y}(t) = \frac{y(t) - \min_t y(t)}{\int [y(t) -
    \min_t y(t)]\, dt},
    \qquad
    \tilde{t} = \frac{t}{\max_t t}.
    \label{eq:normalization}
\end{equation}
This normalization places every GDC component's linear coefficient on a comparable scale, so that each coefficient can be interpreted as a mixing weight in the same sense as a mixture-model proportion.

The number of components $J$ and their initial center parameters $\{\mu_j^{(0)}\}$ are obtained by peak detection on the normalized curve.
Each detected peak seeds the center parameter $\mu_j$ of one component, in the same way that a $k$-means initialization seeds each cluster centroid before iterative refinement.
The growth and decay parameters, $\sigma_j$ and $\lambda_j$, are given a shared small starting value for every component, since no prior information distinguishes one component's shape from another at this stage.

Given these starting values, the nonlinear parameters $\{\mu_j, \sigma_j, \lambda_j\}_{j=1}^J$ are estimated jointly using variable projection with Levenberg-Marquardt \citep{golub1973differentiation}.
Variable projection exploits the fact that, for fixed nonlinear parameters, the linear mixing weights $\{\beta_j\}$ enter the model linearly and can therefore be solved for in closed form at every iteration of the outer nonlinear search, rather than being treated as free parameters alongside $\{\mu_j, \sigma_j, \lambda_j\}$.
This inner linear solve is ridge-regularized rather than computed via the ordinary pseudoinverse, since GDC components with overlapping centers produce a design matrix whose columns are highly correlated:
\begin{equation}
    \hat{\boldsymbol{\beta}} =
    \left(\mathbf{X}^T\mathbf{X} + \lambda_{ridge}\mathbf{I}\right)^{-1}
    \mathbf{X}^T\tilde{\mathbf{y}},
    \label{eq:varpro_ridge}
\end{equation}
where $X_{ij} = \mathrm{GDC}(t_i;\, \mu_j, \sigma_j, \lambda_j)$ is the $j$-th component evaluated at the $i$-th observation, and $\lambda_{ridge} > 0$ is a fixed ridge penalty.
This formulation, following \citet{espanol2023variable}, extends the original Golub-Pereyra variable projection method by replacing its pseudoinverse-based linear step with a Tikhonov-regularized one, so that the inner solve remains well-defined even when the design matrix is ill-conditioned.
Rather than tuning $\lambda_{ridge}$, it is fixed at a small constant for computational simplicity; similar fixed-penalty schemes have been used previously for this purpose \citep{park20071}.
The output of this step is a fixed design matrix, with one column per component, each column equal to that component's normalized GDC density evaluated at the observed time points.

With the nonlinear parameters fixed and the design matrix constructed, the linear mixing weights are estimated.
Because the design matrix's columns remain highly correlated at the fitted nonlinear parameters, ordinary least squares is poorly suited to this final step, so a cross-validated elastic-net penalty is used instead \citep{zou2005regularization, tay2023elastic}.
The elastic-net penalty simultaneously handles the multicollinearity among overlapping components and performs variable selection, shrinking the coefficients of components that are not supported by the data toward zero.
As such, it is preferable to initialize more candidate components than are ultimately needed: a redundant component is removed by the elastic-net step, whereas a component that was never initialized cannot be recovered later. 

\subsection{Comparators and Metrics}
\label{sec:comp_comparators}
GDC is compared against four alternative curve-fitting methods in \textsf{R} \citep{rcore}, chosen to span a range of flexibility and tuning philosophies rather than to represent an exhaustive survey of smoothing techniques. Each method is fit is evaluated on four metrics: root mean squared error (RMSE) and maximum absolute error, which characterize average and worst-case fit quality; Lin's concordance correlation coefficient (CCC) \citep{lawrence1989concordance}, which jointly captures precision and accuracy relative to the observed signal; and the Bayesian information criterion (BIC), which penalizes each method's parameter count directly and is therefore the primary criterion for comparing parsimony rather than fit quality alone.  For methods without a fixed, finite parameter count, effective degrees of freedom (EDF) are substituted for the raw parameter count in both the BIC penalty and the reported parameter total, giving a fairer, more directly comparable measure of complexity against GDC's fixed parametric form. The four methods are as follows:

\begin{itemize}
    \item \textbf{Natural cubic spline.}
    Spline smoothing is among the most widely used nonparametric approaches to curve estimation and imposes no assumptions about the underlying functional form, making it a natural, assumption-free point of reference for GDC's physics-informed parameterization.
    The smoothing parameter, and thus the effective degrees of freedom, is selected internally via generalized cross-validation, so no information from GDC's fit is used in tuning it.

    \item \textbf{Additive smoothing model.}
    Generalized additive models are a standard extension of spline smoothing used broadly across the applied sciences for flexible, data-driven curve fitting, providing a second and methodologically distinct nonparametric comparator to GDC.
    A thin-plate smoothing term is fit with its effective degrees of freedom selected via Restricted Maximum Likelihood, again tuned independently of GDC and by a criterion distinct from the cross-validation used for the spline, so that neither nonparametric baseline is advantaged by sharing a tuning strategy with the other \citep{wood2011fast}.

    \item \textbf{Radial basis expansion.}
    Radial basis function networks are closely related to GDC in structure, both are nonlinear expansions of localized basis components fit by variable projection, so this comparator isolates the contribution of GDC's growth-decay basis itself rather than differences in fitting procedure.
    The number of Gaussian components is fixed to match the number of peaks estimated by GDC, so that the two models are compared at an equal component budget rather than allowing RBF to select its own complexity.

    \item \textbf{Truncated harmonic (Fourier) regression.}
    Harmonic regression is a long-established tool for representing periodic and oscillatory structure and offers a comparator built on an entirely different basis system than GDC's localized components.
    Rather than tuning the number of harmonics by an internal criterion, the smallest number of harmonics needed to match GDC's in-sample fit is selected directly, so the comparison reflects the parameter cost of reaching GDC's accuracy rather than the harmonic model's best attainable fit at unrestricted resolution \citep{ramsay2025fda}.
\end{itemize}

The parameter counts compared across methods are not the same kind of object. 
GDC's count is the number of nonlinear shape parameters and linear mixing weights, an effective rank for a nonlinear model that depends on the fitted solution itself rather than a quantity fixed in advance. 
The effective degrees of freedom for splines and GAMs are a trace-based quantity computed from the linear smoother (hat) matrix. This construct is well defined precisely because those methods are linear in the response conditional on the smoothing parameter, a property GDC's nonlinear basis does not share. 
The FFT's parameter count, by contrast, is a search-selected number of harmonics, chosen post hoc to match GDC's accuracy rather than derived from any smoothing or shrinkage operator.
As such, the complexity and BIC comparisons that follow based on EDF and number of parameters should be treated as approximate, informative of relative model size and the general parsimony-versus-fit trade-off across methods, rather than as an exact equivalence of statistical degrees of freedom.


\section{Simulation Study}
\label{sec:simulation}

Having established the theoretical connection between the GDC functional form and classical stochastic and physical processes, this section evaluates GDC's practical utility on two well-characterized differential equation systems with known parameters of interest.

\subsection{Temporal Analysis of Products}
\label{sec:TAP}

The Temporal Analysis of Products (TAP) reactor provides a further test case for the GDC framework, as its transient response curves arise directly from the solution of transport-reaction ordinary differential equations (ODEs) via Laplace transform methods rather than from an arbitrarily chosen functional form. Under Knudsen diffusion, transport is governed by
\begin{equation}\label{eq:diffusion}
\frac{\partial C}{\partial t} = D \frac{\partial^2 C}{\partial x^2},
\end{equation}
subject to the initial condition $C = C_0$ for $x < 0$ and $C = 0$ for $x > 0$ at $t = 0$, whose solution is the standard diffusion curve (SDC),
\begin{equation}\label{eq:SDC}
    SDC(t, \eta) = N \pi \eta \sum_{n = 0}^\infty (-1)^n (2n + 1) \exp \left( -(2n + 1)^2 \frac{\pi^2}{4} \tau \right),
\end{equation}
where $\eta \, (1/s) = D/\epsilon L^2$, $D$ is the diffusion coefficient, $L$ is length, $\epsilon$ is the bed porosity, and $\tau = t \eta$ is the dimensionless time. This solution is a Gaussian and establishes the direct relation between TAP transport and the GDC framework: \citet{gleaves1997tap} showed that the resulting outlet flux is well approximated by a single GDC term, while related work, \citet{kunz2025statistical}, demonstrated an early form of this correspondence by fitting both Knudsen and non-Knudsen transport with the GDC distribution. 
This simulation extends the viability of the GDC fit by including reaction. When an irreversible surface reaction is superimposed on this transport, the resulting solution is simply the SDC multiplied by an exponential decay, and the corresponding response is well approximated by a single exponentially modified GDC term; the reversible case is likewise derived in closed form by \citet{gleaves1997tap} (Eq.~72), but its solution does not reduce cleanly to a simple combination of GDC terms. This ambiguity motivates using GDC-derived parameters, rather than an assumed mechanism, to probe how well intrinsic kinetic rate constants can be recovered directly from the transient response.

To test this, TAP responses were simulated directly from the governing rate equations. Let the gas-phase concentration be denoted as $A$ and surface concentration $Z$ evolving under the rate $R$ with kinetic coefficients $k_a, k_d, \mbox{ and } k_r$ be the adsorption, desorption and reversibility, respectively. Then, $R_A = k_a A$ for the irreversible case, and $R_A = k_a A - k_d AZ$, $R_{AZ} = R_A - k_r AZ$ for the reversible case, where $k_a$ was held consistent between the two experiments. Irreversible simulations swept $k_a$ over 500 samples from 0.5 to 10, and reversible simulations swept $k_d$ and $k_r$ each over 500 samples from 10 to 30 and 1 to 3, respectively. 
Curve-fitting accuracy was substantially higher for GDC than for the EMG benchmark, with RMSE of 0.002 versus 0.016 for the irreversible case and 0.004 versus 0.024 for the reversible case. GDC also achieved a minimum CCC of 0.9999 (irreversible) and 0.9997 (reversible), alongside a maximum relative RMSE of 0.0017 and 0.0069, respectively, despite the reversible curves being fit with only a single GDC term. 
Because the GDC parameters do not correspond directly to the kinetic rate coefficients of the underlying differential equations, an ordinary least squares model was instead used to relate $1/\exp(\mu)$, $\sigma$, and $\lambda$ to the true rate coefficients across the 500 simulated runs. 
This mapping achieved an $R^2$ of 0.999 for the irreversible coefficient $k_a$, reflecting its near-exact correspondence to the GDC exponential term. For the reversible case, however, $R^2$ values were lower: 0.882, 0.806, and 0.865 for $k_a$, $k_d$, and $k_r$, respectively, consistent with the reversible response requiring curve structure beyond what a single GDC term can fully resolve.
These results indicate that while GDC parameters cannot be interpreted as mechanistic rate constants directly, they can recover intrinsic kinetic information with high fidelity under irreversible kinetics and with reasonable, though reduced, fidelity when reversibility introduces additional unresolved curve structure.


\subsection{Susceptible-Infected-Recovered}
\label{sec:SIR}

\begin{figure}[ht]
  \centering
  \includegraphics[width=8.5cm, height = 6.375cm]{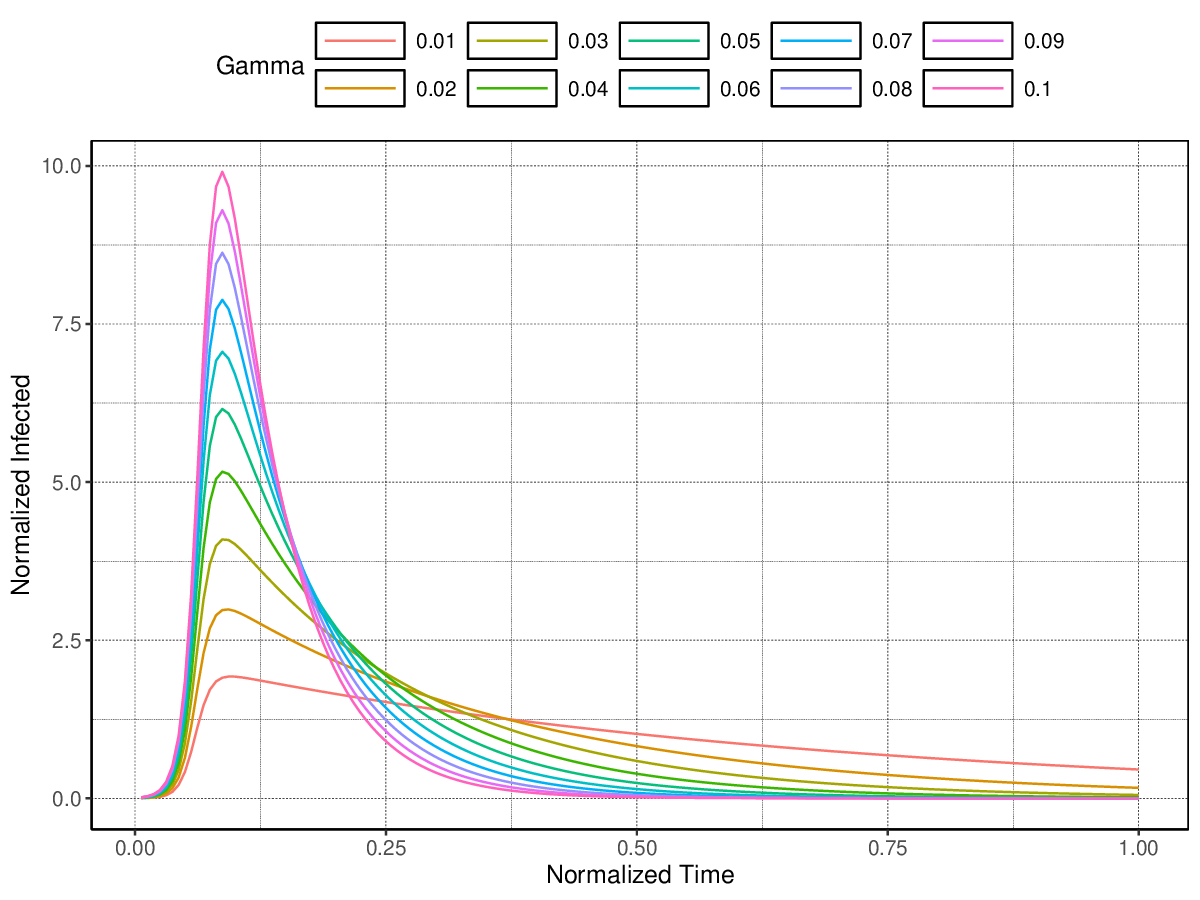}
  \caption{Plots of the infected compartment of the SIR model for a $\beta_S$ of 0.8 and varying values of $\gamma$, obtained directly from solving the differential equations.}
  \label{fig:SIR}
\end{figure}

The Susceptible-Infected-Recovered (SIR) model provides a natural additional test case for the GDC framework, as it describes the temporal evolution of an epidemic through a coupled system of nonlinear ordinary differential equations rather than a simple closed-form curve. 
The population of size $N = S + I + R$ is partitioned into susceptible, infected, and recovered compartments, whose dynamics are governed by
\[
\frac{dS}{dt} = -\frac{\beta_S S I}{N}, \qquad
\frac{dI}{dt} = \frac{\beta_S S I}{N} - \gamma I, \qquad
\frac{dR}{dt} = \gamma I,
\]
where $\beta_S$ is the transmission rate and $\gamma$ is the recovery rate. 
These two parameters jointly determine the basic reproduction number $R_0 = \beta_S/\gamma$, which characterizes the epidemic threshold and overall severity of the outbreak. 
Figure~\ref{fig:SIR} shows the resulting infected-compartment trajectories obtained by numerically solving this system for a fixed $\beta_S = 0.8$ and a range of $\gamma$ values, producing infection curves that vary systematically in peak height and duration. 
Because the infected curve $I(t)$ produced by this system exhibits the same rise-and-decay shape targeted by the GDC formulation, it serves as a suitable candidate for evaluating GDC's ability to recover epidemiologically meaningful quantities directly from the shape of the curve.

To evaluate whether the GDC-derived ratio generalizes across a broad range of epidemic severities, the SIR model was fit across a wide variety of $R_0$ values generated by varying both $\beta_S$ and $\gamma$ jointly. 
The parameter $\beta_S$ was varied from 0.2 to 0.8 in increments of 0.1, and for each value of $\beta_S$, 50 values of $\gamma$ were generated as $\gamma = \beta_S U$, where $U$ was drawn independently from a uniform distribution on $[0.1, 0.9]$. 
Because $R_0 = \beta_S/\gamma = 1/U$ depends only on the drawn proportion and not on the absolute value of $\beta_S$, different values of $\beta_S$ can produce similar or overlapping $R_0$ values by chance.
Fitting the GDC model to the numerically solved infected curves yields a minimum CCC of 0.9 between the ODE solution and the GDC estimate, indicating that the GDC functional form closely reproduces the SIR infection dynamics across the full grid of $\beta_S$ and $\gamma$ values considered.
To connect the GDC parameters to the epidemiological quantity of interest, the inverse of $\Pi_\tau$, $1/\Pi_\tau = \lambda/\exp(\mu)$, is compared against the true $R_0 = \beta_S/\gamma$; a linear model relating the two quantities, fit on the noise-free ODE solutions, $R_0 = 1.562 + 7.008\,(1/\Pi_\tau)$, achieves an $R^2$ of 0.947 and a mean squared error (MSE) of 0.151, demonstrating a moderately strong linear correspondence between the GDC-derived ratio and the true basic reproduction number.

To assess how this correspondence holds under noisy observations, Gaussian noise was added directly to the ODE-solved infected curves, prior to GDC fitting, across a grid of five variances ranging from $1e{-4}$ to 0.25. At each noise level, 500 values of $\gamma$ were drawn from the same range used in the noise-free fit, with $\beta_S$ held fixed throughout at a value of 0.8. The resulting MSE between the true and GDC-estimated $R_0$ was recorded alongside the ratio $\sigma^2/MSE$, as summarized in Table~\ref{tab:sir_noise}. 
Even the smallest noise variance considered produces an MSE roughly two orders of magnitude larger than the noise-free baseline, and this gap continues to widen as $\sigma^2$ increases to 0.25. However, the comparatively stable ratio $\sigma^2/MSE$ across the grid indicates that, beyond this initial jump, additional error scales approximately proportionally with the amount of noise added to the ODE solution.
\begin{table}[ht]
    \centering
    \begin{tabular}{c| c c c c c}
        $\sigma^2$ & 0.0001  &0.0176 & 0.065 & 0.143 & 0.250  \\
        \hline 
        $MSE$ & 0.009&1.177 &5.289 &11.612 & 25.658 \\
        $\sigma^2 / MSE$ & 0.012 & 0.015 & 0.013 &0.012 & 0.009
    \end{tabular}
    \caption{MSE between the true and GDC-estimated $R_0$ across a grid of Gaussian noise variances $\sigma^2$ applied to the SIR curves, with 500 samples of $\gamma$ drawn at each noise level. The ratio $\sigma^2/MSE$ remains approximately constant across the grid, indicating that estimation error scales proportionally with the level of noise in the data.}
    \label{tab:sir_noise}
\end{table}

\section{Empirical Validation}
\label{sec:results}

The four applications to measured data that follow are organized to build in complexity rather than to repeat a fixed template.
Each subsection accordingly differs in which comparators and diagnostics it emphasizes, reflecting what is most informative for that particular application rather than a uniform checklist applied regardless of fit.

\subsection{Housing Prices: Nonnegative Economic Response}
\label{sec:emp_housing}
\begin{figure}[ht]
  \centering
  \includegraphics[width=8.5cm, height = 6.375cm]{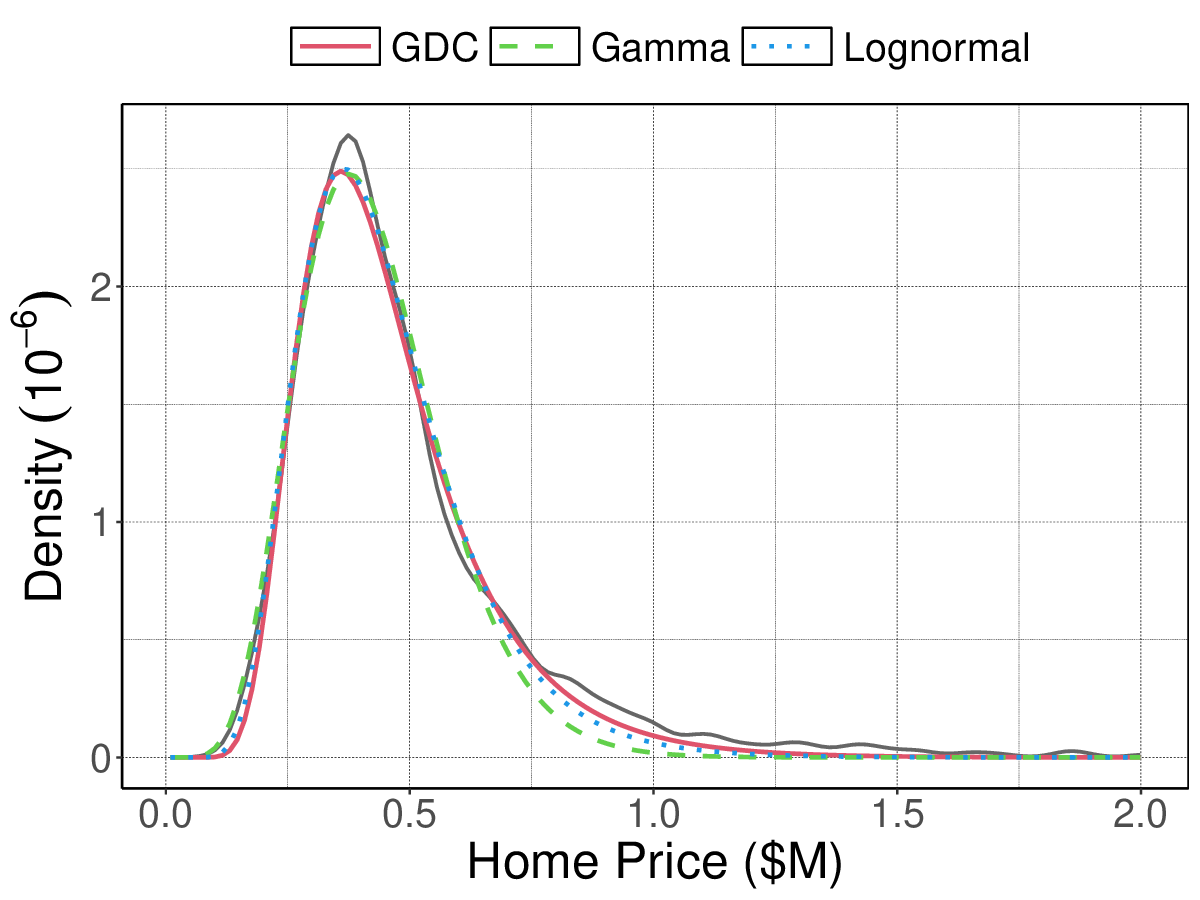}
  \caption{Comparison of Virginia home prices (black) to distribution estimates.}
  \label{fig:homeprices}
\end{figure}

As a first, more pedagogical example of fitting the GDC, we consider housing prices. 
The data were taken from housing sales in Albemarle County, Virginia, where the original objective of the exercise was to predict housing prices using Gamma regression with a log link \citep{ford2020getting}. 
Rather than examining the effects of predictors such as square footage, we instead compare the fit of the GDC against two more traditional distributions, the Gamma and the lognormal. 
Although the majority of this paper has focused on the physical interpretation of the basis functions, this example still satisfies two conditions relevant to that motivation: prices are constrained to be nonnegative, and housing prices are widely understood to arise from a multiplicative rather than additive process, consistent with the lognormal-type generative mechanism developed earlier in the paper.
Figure~\ref{fig:homeprices} compares the density estimate of the home prices for the GDC to the Gamma and lognormal distributions. 
Examining the BIC for each method yields values of -426.9, -512.5, and -562.7 for the Gamma, lognormal, and GDC fits, respectively, with the more negative value indicating the better fit. 
Lin's CCC tells a consistent story, at 0.991, 0.995, and 0.997 for the Gamma, lognormal, and GDC fits, respectively. 
Together, these metrics show that the GDC outperforms both traditional distributions despite its larger number of parameters.
However, in practice, any of the three may work reasonably from a purely qualitative standpoint.

Perhaps more interesting is the interpretation of the GDC in this context.
Because the covariate here is price rather than time, $\mu$ and $\lambda$ are more precisely read as characteristic price scales rather than timescales, though the same structural interpretation applies: the lognormal CDF term governs the price range over which the density rises, and the exponential term governs the price range over which it subsequently thins. Here, $\exp(\mu) = 343140.50 \pm 1078.97$ is the growth-phase price scale, $\lambda = 165025.10 \pm 3452.76$ is the decay-phase price scale, and $\sigma = 0.33 \pm 0.01$ describes the relative shape of the lognormal.
Because $\exp(\mu) > \lambda$, the growth phase operates over a substantially wider price range than the decay phase, indicating that the bulk of the price distribution's rise is spread across a wider dollar range than the thinning of its upper tail; the gap between the two, $\exp(\mu) - \lambda \approx \$178{,}115$, quantifies this directly in dollar terms.
The ratio $\Pi_\tau = \exp(\mu)/\lambda \approx 2.08 \pm 0.04$ quantifies this asymmetry for this housing market and provides a basis for comparison across different regions: a growth-phase price scale roughly double the decay-phase price scale.

\subsection{Solar Irradiance: Spectral Decomposition}
\label{sec:emp_solar}
This next example considers prediction of the solar spectral irradiance measured under the American Society for Testing and Materials (ASTM) E-490 standard \citep{national20002000}.
The spectral irradiance functional profile has been established through Planck's law:
\begin{equation*}
    E(\lambda_p, T) = \frac{2 \pi h c^2}{\lambda_p^5} \left(e^{h c / (\lambda_p k_B T)}-1\right)^{-1}
\end{equation*}
where $\lambda_p$ is a set of wavelengths, $T$ is temperature in Kelvin, $h$ is Planck's constant, $c$ is the speed of light, and $k_B$ is Boltzmann's constant.
Wien's displacement constant, $b$, can be derived directly from $h$, $c$, and $k_B$ rather than taken as a separately tabulated value, giving $b \approx 2897.77\ \mu\text{m}\cdot\text{K}$ \citep{reif2009fundamentals}.
At $T = 5777\,$K,  the constant derived above predicts a peak wavelength of $\lambda_{peak} = b/T \approx 0.5016\ \mu\text{m}$.
Consider the two different terms of Planck's law: the power-law prefactor decreases monotonically with wavelength, while the occupation factor increases monotonically with wavelength.
This resembles the form of the GDC but leverages a power law for the decay, rather than an exponential, and an unbounded growth term for the increase, rather than the saturating lognormal CDF.

\begin{figure}[ht]
  \centering
  \includegraphics[width=8.5cm, height = 6.375cm]{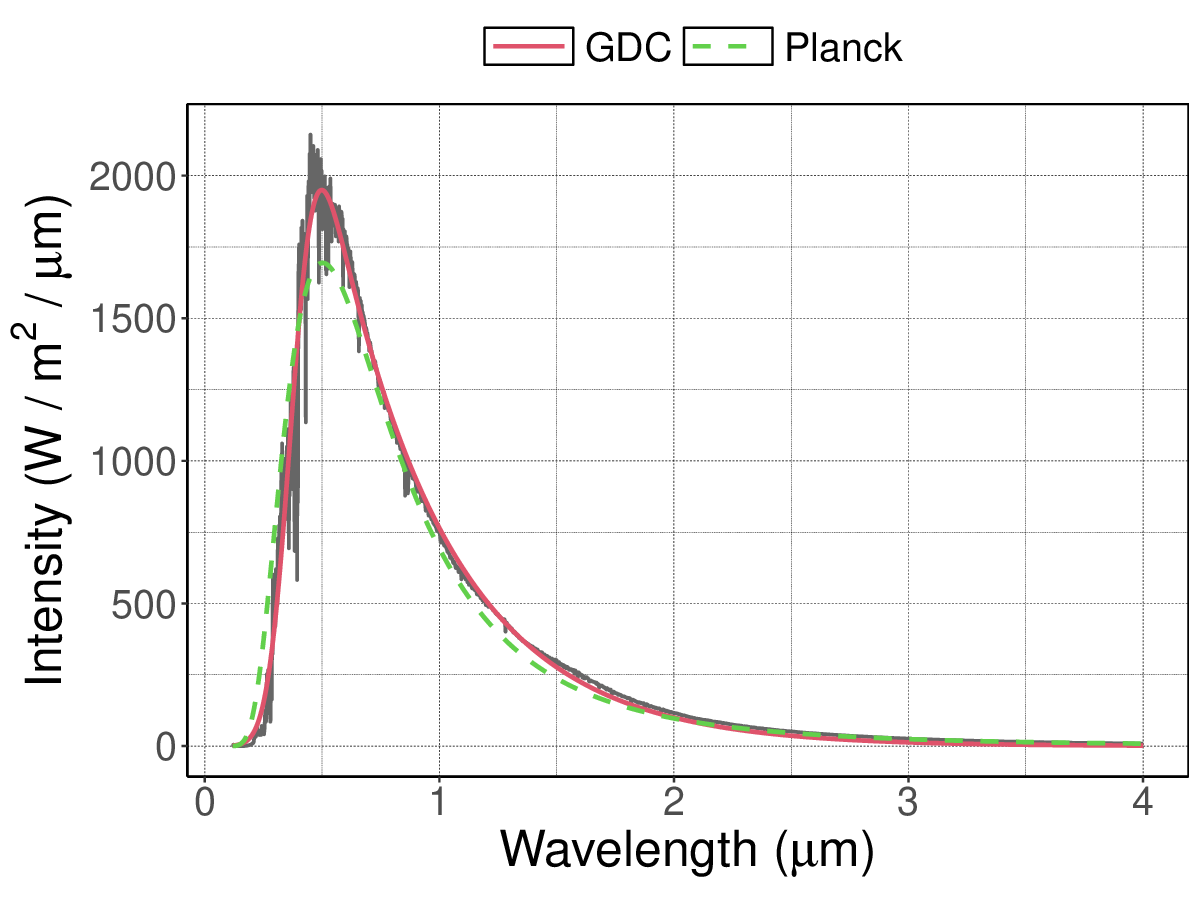}
  \caption{Fit of solar irradiance measurements compared to Planck's law.}
  \label{fig:planck}
\end{figure}

Figure~\ref{fig:planck} compares GDC against Planck's law fit to the observed solar spectral irradiance data where $(\exp(\mu), \sigma, \lambda) = (0.388 \pm 0.002, 0.225 \pm 0.005, 0.476 \pm 0.007)$.
The GDC's fitted peak location of $0.503$ sits within approximately $0.3\%$ of this first-principles value, indicating that the fit preserves the correct physical scaling between temperature and peak wavelength even though that constraint was never imposed directly.
Additionally, the overall fit went from a RMSE of 141.738 using Planck's Law to 76.611 using the GDC, a decrease of approximately $54\%$.
The result is not that GDC improves on Planck's law as physics, since Planck's law is derived from first principles.
Rather, it demonstrates that GDC's added flexibility recovers the correct physical peak location while being based on a set of physically understood functions.

\subsection{COVID-19 Incidence: Epidemic Wave Decomposition}
\label{sec:emp_covid}

This demonstration can be considered an extension of the SIR model in Section~\ref{sec:simulation}, applied instead to COVID-19 data.
Specifically, Florida's daily infection counts from the NY-Times COVID-19 dataset were used to validate the fitted curve \citep{nyt2021covid}.
This case study goes beyond estimation of a single distribution, since the data represent multiple successive infection waves with potential overlap between them.
Rather than comparing GDC against the SIR model, as in Section~\ref{sec:simulation}, this demonstration showcases GDC's ability to match the fit quality of RBF, GAM, FFT, and spline regression while being more parsimonious, i.e., using substantially fewer parameters.

\begin{figure}[ht]
  \centering
  \includegraphics[width=8.5cm, height=6.375cm]{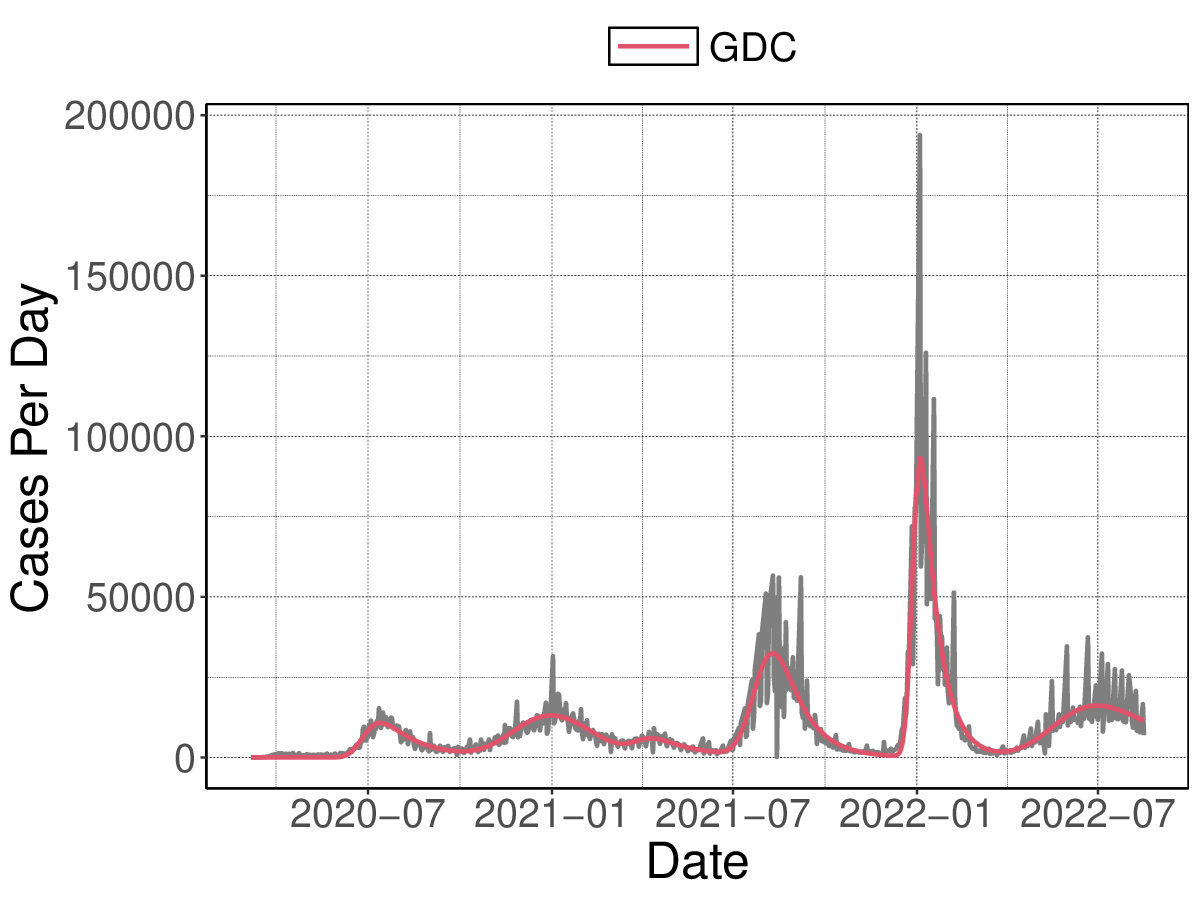}
  \caption{Fit of COVID-19 cases per day for Florida.}
  \label{fig:covid}
\end{figure}

Figure~\ref{fig:covid} displays the fitted curve for GDC against the observed daily case counts.
Each wave produces an asymmetric, unimodal incidence curve consistent with compartmental epidemic dynamics; GDC captures both the asymmetric rise and the exponential decay of each wave.

Table~\ref{tab:COVID} reports the fit comparison under a revised, more tightly matched set of fitting procedures.
GDC was fit using six estimated peaks, yielding 24 parameters, and RBF was fit using the same VarPro procedure with the same number of estimated peaks, yielding 18 parameters.
Spline was fit using R's internal methods for selecting the appropriate number of degrees of freedom, and GAM was initialized at the rounded degrees of freedom obtained from that Spline fit (set as $k$) before REML estimated the total effective degrees of freedom.
FFT was fit by selecting the number of Fourier harmonics that most closely matched GDC's performance without overfitting, since the goal here is to compare the number of parameters each method needs to reach comparable accuracy, not to find FFT's best possible fit.

Under this matched-function comparison, GDC and FFT are nearly tied on RMSE (0.648 vs. 0.647) and CCC (0.853 vs. 0.852), with GDC reaching this level of accuracy using roughly a third of FFT's parameter count (24 vs. 68).
Spline and GAM both trail GDC on every fit-quality metric despite using more than twice as many EDFs (65.127 and 53.139, respectively, vs. 24).
RBF trails all four other methods on RMSE (0.694) and CCC (0.830), and produces the highest maximum error (11.136), though by a narrower margin than the other comparators. This is consistent with a symmetric Gaussian basis being less well suited than an asymmetric one to the rise and exponential decay of each wave.
GDC achieves the lowest maximum error of any method, indicating that its parsimony is not achieved by sacrificing fit at the waves' peaks.

The BIC column makes the parsimony argument directly rather than leaving it to be inferred from the parameter counts alone.
GDC's BIC ($-510.612$) is the lowest of the five methods, followed by RBF ($-443.399$), with both well ahead of FFT ($-230.931$), Spline ($-231.587$), and GAM ($-211.774$); RBF's comparatively strong BIC, despite its weaker RMSE and CCC, reflects its smaller parameter count (18) rather than a competitive fit, underscoring that BIC and raw fit quality can favor different methods depending on how heavily complexity is penalized.
Because BIC penalizes parameter count directly, this gap is best treated as suggestive rather than decisive, but it is nonetheless the most favorable of the four comparisons to GDC on a parsimony basis.

Two additional checks confirm that GDC's advantage is not an artifact of under-tuning the competing methods.
First, extending FFT to its maximum resolution, 385 harmonics (half the length of the signal), drops RMSE to 0.388, but this configuration is highly overfit and is not a meaningful point of comparison.
This is reported only to confirm that FFT's matched-accuracy parameter count (68) was not simply an artificially low ceiling.
Second, GAM's minimum achievable RMSE at its maximum estimated EDFs (62.84) is 0.652, which still does not surpass GDC's 0.648, indicating that additional GAM flexibility does not close the gap with GDC even at its practical upper limit.
\begin{table}
  \centering
  \begin{tabular}{l | cc  ccc}
    Method  & Terms / EDF & RMSE & Max Error &BIC & CCC\\
    \hline
    GDC    & 24 & 0.648 & 9.671 & -510.612& 0.853\\
    RBF    & 18 & 0.694 & 11.136 &-443.399& 0.830\\
    Spline & 65.127 & 0.650 & 10.778 & -231.587& 0.840\\
    GAM    & 53.139 & 0.659 & 11.092 & -211.774& 0.835\\
    FFT    & 68 & 0.647 & 10.281 & -230.931& 0.852
  \end{tabular}
  \caption{Fit comparison across methods for the Florida COVID-19 waves.}
  \label{tab:COVID}
\end{table}
Table~\ref{tab:COVIDinterpretation} examines the fitted GDC parameters across the six waves.
Following Section~\ref{sec:simulation}, $R_0$ is estimated directly from $\Pi_\tau$, the ratio of the decay and center terms, and the lognormal CV of the growth phase is reported alongside it.
The $R_0$ values decline fairly steadily across the six waves except for the beginning in early 2022, where $R_0$ rises to 2.122 from the preceding wave's 1.739, which may indicate the highly transmittable Omnicron variant in late 2026 \citep{tian2022emergence}.
The CV values are below 1 for every wave, indicating that the growth phase in each case is sub-exponential and closer to a step-change onset than a long-tailed one.
The final wave (2022-06) is the exception worth noting: its CV of 0.245 is roughly 2--6 times higher than any other wave's, suggesting a comparatively more gradual, drawn-out onset.
However, this could be due to incomplete information (i.e., data cut off prior to the curve's completion) used in estimating the final GDC curve.

\begin{table}
  \centering
  \begin{tabular}{l | c c c c c c}
    Date  & 2020-07 & 2021-01 & 2021-04 & 2021-08 & 2022-01 & 2022-06\\
    \hline
    $R_0$ & 3.558   & 2.348   & 1.905   & 1.739   & 2.122   & 1.576\\
    CV    & 0.105   & 0.036   & 0.030   & 0.009   & 0.042   & 0.245\\
  \end{tabular}
  \caption{$R_0$ and lognormal growth-phase CV for each GDC-fitted COVID-19 wave.}
  \label{tab:COVIDinterpretation}
\end{table}

\subsection{Sunspot Activity: Solar Cycle Decomposition}
\label{sec:emp_sunspot}

\begin{figure}[ht]
  \centering
  \includegraphics[width=8.5cm, height=6.375cm]{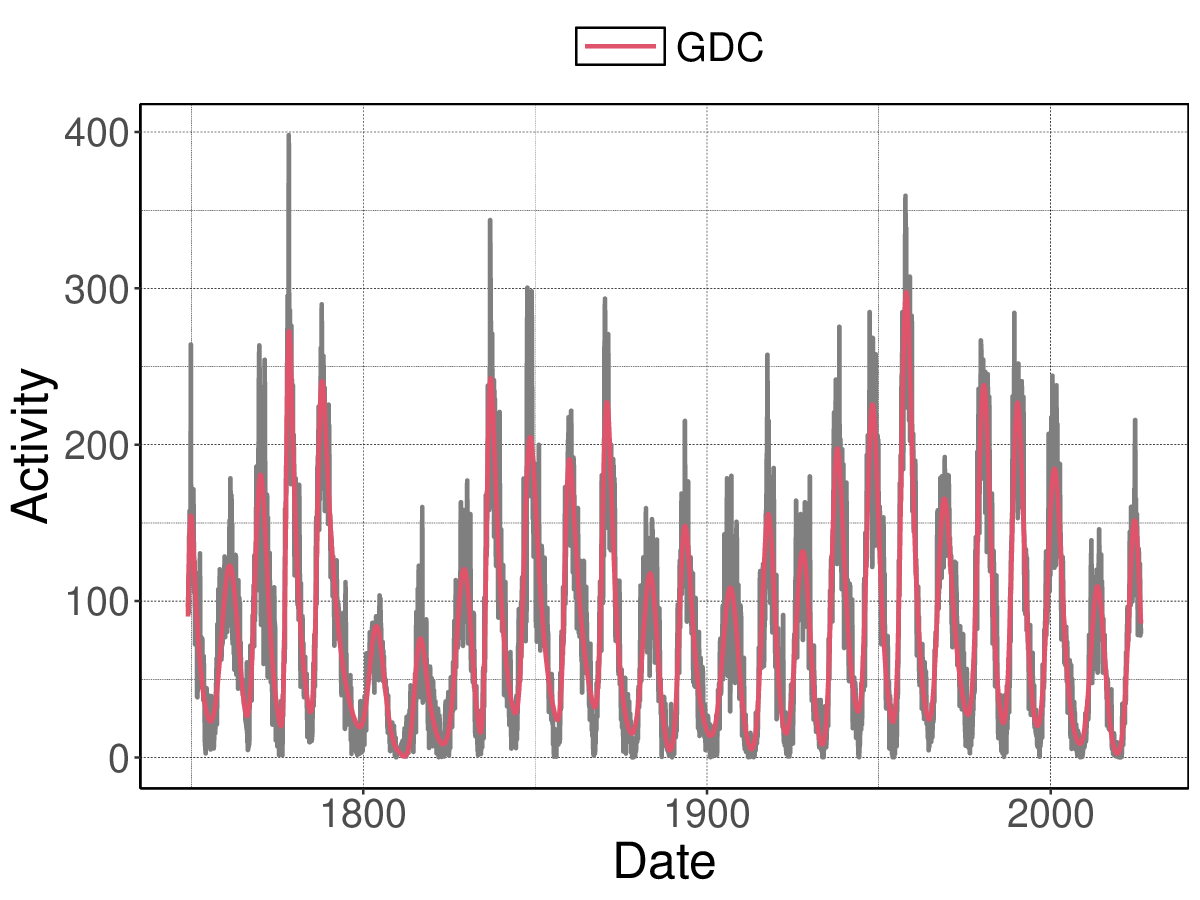}
  \caption{Fit of solar flares.}
  \label{fig:solarflaresfit}
\end{figure}

Monthly sunspot counts spanning over 150 years exhibit approximately 11-year cycles with an asymmetric rise and fall \citep{noaa2026solarcycle}.
Accurately characterizing these cycles has practical importance beyond solar physics itself, since solar activity drives space weather phenomena that affect satellite operations, high-frequency radio communication, power grid stability, and the reliability of GPS and other satellite-based positioning systems \citep{chen2024solar}.
Visual inspection of the full 150-year record identifies approximately 27 distinct peaks of sunspot activity, and Figure~\ref{fig:solarflaresfit} shows the corresponding GDC fit overlaid on the observed monthly counts across the entire series.
The objective of this analysis is to estimate the characteristic rise and fall timescales of each cycle, which can then be used to anticipate the timing and shape of upcoming cycles.

First, Table~\ref{tab:solarflares} compares GDC against spline, GAM, and FFT regression on this dataset.
Because sunspot cycles are approximately, though not perfectly, periodic, FFT is a natural candidate, and its performance below reflects that.
GDC's RMSE (0.327) and CCC (0.921) are close to, though slightly behind, the other three methods, each of which lies within a narrow band (RMSE 0.307--0.319, CCC 0.927--0.931).
GDC also produces the lowest maximum error (1.694) of any method, meaning that despite the marginally higher average RMSE, GDC is not making larger localized errors at any individual point in the series.
GDC requires only 104 parameters, roughly 30--45\% fewer than spline (183.185), GAM (167.722), or FFT (148), and this parsimony is reflected directly in BIC.
GDC's BIC (-6450.240) is the lowest, and therefore the best, of the four methods, despite its RMSE and CCC being marginally worse.
This indicates that GDC's small, consistent fit-quality gap is more than offset by its substantially lower model complexity, making it the preferred model under a complexity-penalized criterion even though it is not the single best fit by RMSE alone.

\begin{table}[ht]
  \centering
  \begin{tabular}{l | cc c ccc}
    Method  & Terms / EDF & RMSE & Max Error& BIC & CCC\\
    \hline
    GDC    & 104     & 0.327 & 1.694 & -6450.240 & 0.921\\
    Spline & 183.185 & 0.307 & 2.088 & -6212.243 & 0.931\\
    GAM    & 167.722 & 0.310 & 2.100 & -6167.755 & 0.930\\
    FFT    & 148     & 0.319 & 2.121 & -6262.771 & 0.927\\
  \end{tabular}
  \caption{Fit comparison across methods for solar flares.}
  \label{tab:solarflares}
\end{table}

\begin{figure}[ht]
  \centering
  \includegraphics[width=8.5cm, height=6.375cm]{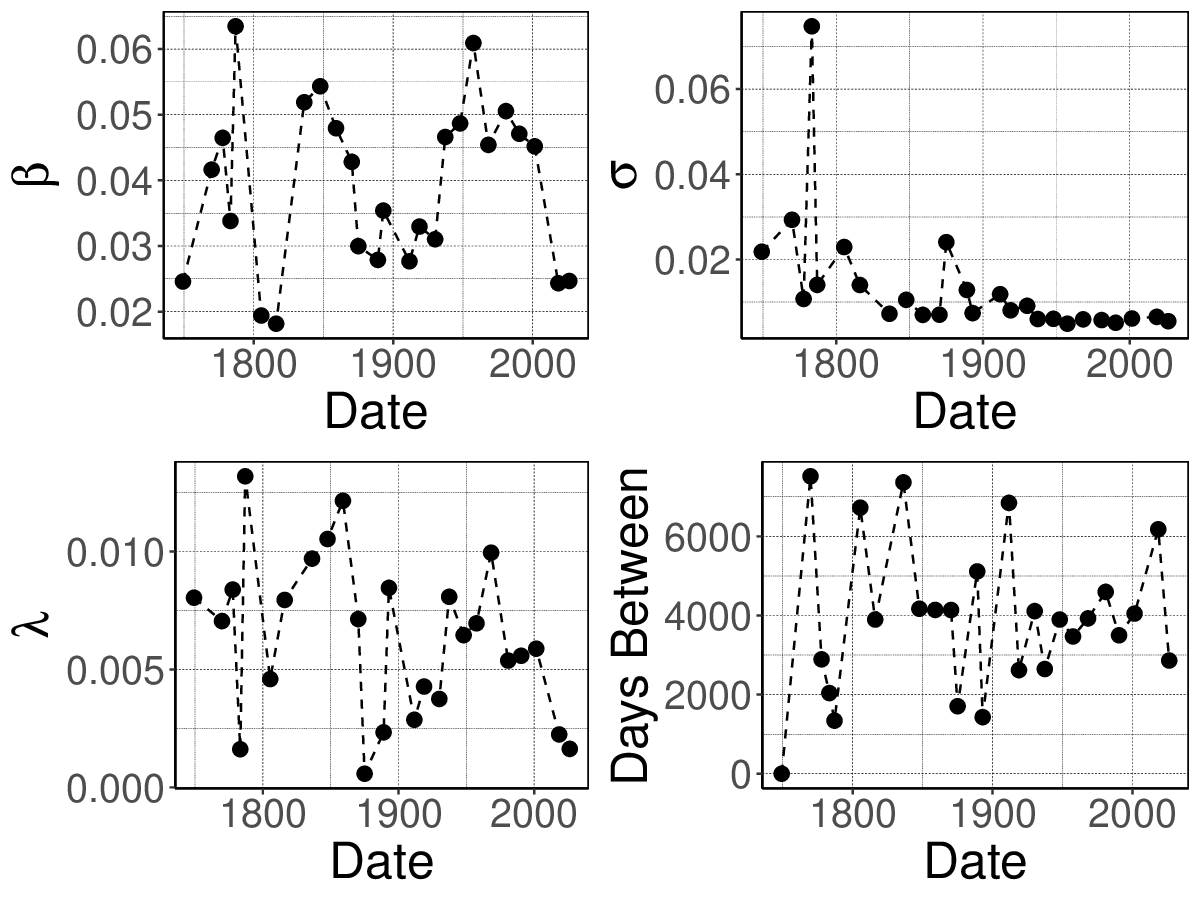}
  \caption{Coefficients of the solar flare events.}
  \label{fig:solarflarescoefs}
\end{figure}

Figure~\ref{fig:solarflarescoefs} shows the fitted GDC coefficients across all 27 identified cycles, revealing structure beyond the raw fit statistics in Table~\ref{tab:solarflares}.
The magnitude $(\beta)$ coefficient follows an approximately cyclic pattern rather than a stable or trending one, suggesting a longer-period modulation of cycle intensity on top of the roughly 11-year spacing.
The $\sigma$ value trends toward a stable value, with a median of approximately 0.006 over the most recent 10 cycles, while the location $(\exp(\mu))$ difference between days has a median spacing of about 3927 days (10.75 years).
Together, these trends indicate a decay timescale that is stabilizing alongside a growth timescale that continues to set the dominant periodicity which can be used to forecast the timing and shape of upcoming cycles.

\section{Conclusion}
\label{sec:conclusion}

The central result of this paper is that a basis function's parameters can be made to carry physical meaning without giving up the fit quality expected of a general-purpose smoother, provided the basis is derived from the same governing equations that produced the data.
The empirical results are evidence that this trade need not be as steep as it is often assumed to be: a basis built for interpretability can still be competitive on fit quality, which suggests that physically motivated bases deserve a place alongside splines, GAMs, and other general-purpose smoothers as a default option, not only as a specialized tool reserved for settings where the governing equation is already known with certainty.

Beyond parsimony, the GDC basis offers an interpretability advantage rooted in how its parameters are read.
Because each component's parameters are timescales rather than abstract shape statistics, a fitted GDC curve can be read the way a differential equation is read, in terms of rates of growth and decay, rather than in terms of moments of a fitted density.
This makes quantities such as the dimensionless ratio $\Pi_\tau$ and the CV directly meaningful from the fitted parameters, without a separate post hoc analysis step to connect the fit back to the underlying dynamics.

The current approach also has limitations.
The number of components and their initial center locations are determined by peak detection on the normalized response prior to fitting, a role analogous to selecting the number of clusters and their initial centroids before running $k$-means. This initialization step carries the same practical caveat as $k$-means: closely spaced or weak peaks may be merged or missed, particularly in signals with substantial overlap between components.
The ridge penalty used to stabilize the variable projection step is similarly fixed rather than jointly optimized with the nonlinear parameters, which is a reasonable practical simplification but not a fully principled treatment of the regularization strength.

Several extensions are natural directions for future work, though we do not pursue them here.
Connections to dynamic mode decomposition may offer a route to extending GDC from single-curve fitting to multi-channel or spatiotemporal data, where a shared set of growth-decay modes could be estimated jointly across channels\citep{brunton2016discovering}.
Also, embedding GDC components within a time-series framework, rather than fitting a single static curve, could allow the growth and decay timescales themselves to evolve, offering a physically motivated alternative to standard time-varying-coefficient models.

    \section{Acknowledgments}
    Work supported through the INL Laboratory Directed Research $\&$ Development (LDRD) Program under DOE Idaho Operations Office Contract DE-AC07-05ID14517.
    
    The authors of this work used Claude Sonnet 5.0 in the preparation of this document for general brainstorming and to improve grammar, sentence structure, and transitions. The authors have reviewed and take full responsibility for the resulting content.

\bibliography{references}

\end{document}